\documentstyle[aps]{revtex}
\begin{document}
\tightenlines

\title{Geometrical approach to light  in inhomogeneous media}

\author{P.\ Piwnicki\thanks{e-mail: piw@quantopt.kth.se}}
\address{Department of Laser Physics and Quantum Optics,
 SCFAB\\
Royal Institute of Technology (KTH), 10691 Stockholm, Sweden}
\maketitle
\begin{abstract}
Electromagnetism in an inhomogeneous dielectric medium 
at rest is described using the  methods of differential geometry. 
In contrast to a general relativistic approach the electromagnetic 
fields are discussed in  three-dimensional space only.
 The introduction of an appropriately 
chosen three-dimensional metric leads to a significant 
simplification of the description of light propagation 
in an inhomogeneous medium: light rays become geodesics 
of the metric and the field vectors are parallel transported 
along the rays. The new metric is connected to the usual 
flat space metric diag[1,1,1] via a conformal transformation 
leading to new, effective values of the medium parameters 
$\tilde{\varepsilon}$ and $\tilde{\mu}$ with 
$\tilde{\varepsilon}\tilde{\mu}=1$. The corresponding 
index of refraction is thus constant and so is the effective 
velocity of light.  Space becomes effectively 
empty but curved. All deviations from straight-line propagation 
are now due to  curvature. The approach is finally 
used for a discussion of the Riemann-Silberstein vector, an
alternative, complex formulation of the electromagnetic fields. 
\end{abstract}
\date{today}

\section{Introduction}
During the past few years much research has been done in the field of
analog laboratory models of general relativity. A particularly
fruitful approach was the discussion of the
propagation of signals in moving media. In an inhomogeneously moving
medium, light \cite{movmed1} or sound \cite{sound} signals deviate from a
straight path creating the impression of an attractive force drawing
them towards some point.  This effect has led to the notion of optical
and acoustical black holes, velocity profiles where all signals
coming sufficiently
 close to some point are so strongly attracted that they fall
into it. From the theoretical point of view 
the most intriguing  result of these
investigations was the realization that these effects can actually be
described in geometrical terms: an effective space-time metric
creates a space-time curvature and thus leads to the deviation from
the straight path. Although the idea of creating table-top experiments
being models of astronomical objects triggered a particular interest
in the physics of moving homogeneous media, they are by no means
the only systems where differential geometry can be applied to describe
the propagation of light in a dielectric. 

The system described in the present paper is an
inhomogeneous dielectric at rest, i.e., a dielectric with 
electric permittivity
 $\varepsilon$ and  magnetic permeability $\mu$ depending on
position. The best known everyday example of such a medium is the 
air above a hot road which reflects light coming from the sky. Here
the effect is due to the different temperatures in adjecent layers
of air which lead to a variable density. A more extreme example is
the interface between air and glass where the index of refraction
$n=\sqrt{\mu\varepsilon}$ varies discontinuously, but this kind of system will
not be considered here. 

The propagation of a light ray in an inhomogeneous medium has usually
been described in terms of Fermat's principle:  light moves
 along  paths with the shortest optical length, i.e., the paths
that give the shortest transit time. The fact that  light rays 
extremize some measure of length suggests that they might be geodesics
of the corresponding metric. That this actually is the case with a
metric of the form
 $ds^2=n^2(dx^2+dy^2+dz^2)$ has been shown by Bortolotti in his
1926 paper \cite{bort}. The propagation of the field vectors along the
ray has been addressed for the first time by Rytov \cite{rytov} who
discussed this topic in terms of classical differential geometry,
calculating the rotation of the field vectors in relation to the
principal normal of the trajectory.  
Kline and Kay show in their book 
\cite{klinekay} that the field vectors are actually parallel
transported along the curve in the sense of Levi-Civita
parallelism. Unfortunately, the authors stick to a mathematical
notation as used in the differential geometry of curves which makes the
calculations rather cumbersome and conceals the physical meaning of
the results. 
Finally, Solimeno, Crosignani, and DiPorto \cite{scd} mention that it
is possible to introduce a metric tensor and identify the rays as geodesics
in this metric. Furthermore, they represent the ray equation in the form
of a geodesic equation in terms
of Christoffel symbols. They do not, however, describe the fields and
the field vectors in this formalism.

The aim of the present paper is to give an account of how
electrodynamics in an inhomogeneous medium can be represented in an
alternative metric. We will extend the discussion beyond the ray
equation discussed in \cite{scd}. 
With a few exceptions all calculations will be
performed in the convenient notation of the Ricci calculus as used in
the general theory of relativity. This keeps the calculations simple,
gives them a clear meaning and makes them easily accessible to anyone
acquainted with the  methods of general relativity.  We will confine
our attention to metrics connected to the standard metric by a
conformal transformation, i.e., they differ by a scalar factor.  Some
of the calculations are valid in the present form for a more general
class  of metrics, but as these metrics are of no
physical significance for the systems considered here they will not be
discussed. The more
general system -- the anisotropic medium -- requires still more sophisticated
mathematical methods. 

The paper is built up as follows: 
In sec.\ \ref{equations} 
we introduce the general conformally transformed metric and
show how the fields $\mu$, $\varepsilon$ and the wave equation are
transformed under a conformal transformation of the metric. In sec.\
 \ref{geomoptics}
the approximation of geometrical optics is introduced and the wave
equation is rewritten as a series in powers of  the inverse wave number. 
Imposing on the lowest order term the condition that the trajectory
has to be a geodesic leads to the metric introduced by Bortolotti.
In sec.\ \ref{partrans} the
propagation of the field vectors is discussed together with the
intensity flow. The wave equation in the new metric is given. Finally
in sec.\ \ref{rssec} the Riemann-Silberstein (RS) vector in an inhomogeneous
dielectric is discussed. This concept was introduced by Riemann
 \cite{weber} and its main
properties have been discussed by Silberstein \cite{silberstein}. Using
the
name Riemann-Silberstein vector, we follow
Bia{\l}ynicki-Birula \cite{birula}.  The RS-field allows for an alternative,
complex, formulation of electrodynamics, and is  considered a
candidate for the photon wave function \cite{birula}.

\section{Field equations}\label{equations}
In the present section we consider the electromagnetic field
equations in an
inhomogeneous dielectric medium.
 As usual the electric and magnetic fields in this system can be
described by Maxwell's equations
\begin{eqnarray}\label{maxwell3d}
\nabla \cdot {\bf B}=0,\qquad \nabla\times {\bf
  E}=-\frac{\partial {\bf B}}{\partial t},\\ \nonumber
\nabla \cdot {\bf D}=0,\qquad \nabla\times {\bf
  H}=\frac{\partial {\bf D}}{\partial t}.
\end{eqnarray}
The properties of the medium are concealed in the connection between
the  fields, which in our case can be written in the form of
the constitutive equations:
\begin{equation}\label{const3d}
{\bf D} =\varepsilon\varepsilon_0 {\bf E}, \qquad {\bf B}
={\mu\mu_0}{\bf H}
\end{equation}
where we allow the electric permittivity and the magnetic
permeability to depend on
position. Maxwell's equations (\ref{maxwell3d}) and the constitutive
equations (\ref{const3d}) allow us to derive the wave equation in the
medium and, consequently,  to discuss the propagation of light.
Unfortunately, in general  the equation  obtained in this way 
includes many terms which do not have a clear physical
interpretation. In this section we  show that the
introduction of an alternative space structure -- a changed metric -- can
simplify the equations considerably and give them a clear physical
meaning. 

A metric is required, whenever one wishes to define
 lengths of vectors. The length of a vector $\bf a$ can then
be written in the form
\begin{equation}\label{ametric}
|{\bf a}|=\sqrt{ a^\mu a^\nu \gamma_{\mu\nu}}
\end{equation}
where $\gamma_{\mu\nu}$ is the metric tensor.  For the reader familiar
  with
  general relativity we should emphasize that the
  calculations are performed in a three-dimensional space and that the
  metric tensor of flat space is the unit matrix. In (\ref{ametric}) 
  we assume Einstein's summation convention, dropping the summation symbols
and implicitly summing over repeated indices. 
 In a space equipped with a nontrivial metric tensor,   two different
versions of the same vector, the co- and the contravariant one, have to
be considered. They are mapped into each other by the metric tensor
and its inverse. These tensors are said to raise and lower the vector's
indices because  the co- and contravariant vectors are denoted by 
lower and 
 upper indices, respectively:
\begin{equation}\label{updown}
 a_\mu =a^\nu \gamma_{\mu\nu}.
\end{equation}
In ordinary three-dimensional electrodynamics there is
no need for an 
explicit introduction of the metric tensor
which is simply the identity matrix, and the 
  components of the
co- and contravariant vectors are equal.
In the present paper, however, we will allow for more general
metric tensors which correspond to  a more general form of Maxwell's
equations. When Maxwell's equations are written in coordinates, the antisymmetry
of the curl operator is represented by the antisymmetric tensor
$e^{\alpha\beta\gamma}$. In the case of general metrics this has to be
replaced by the generalized form:
\begin{equation}
\eta^{\alpha\beta\gamma} =\frac{1}{\sqrt{\gamma}}e^{\alpha\beta\gamma}
\end{equation}
with
\begin{equation}
\gamma=\mbox{det}(\gamma_{\mu\nu}).
\end{equation}
The covariant metric tensor $\gamma_{\mu\nu}$ is the inverse of 
$\gamma^{\mu\nu}$:
\begin{equation}
\gamma_{\nu\mu}\gamma^{\mu\rho}=\delta^\rho_\nu.
\end{equation}
Following the procedure used by Landau and Lifshitz
\cite{ll2100ex} one can
rewrite Maxwell's equations in the form
\begin{equation}\label{maxwell3d1}
\eta^{\alpha\beta\gamma}\nabla_\beta E_\gamma+\frac{\partial
  B^\alpha}{\partial t}=0,\qquad
 \eta^{\alpha\beta\gamma}\nabla_\beta H_\gamma-\frac{\partial
  D^\alpha}{\partial t}=0
\end{equation}
and
\begin{equation}\label{maxwell3d2}
\nabla_\alpha D^\alpha=0,\qquad\nabla_\alpha B^\alpha=0.
\end{equation}
The symbol $\nabla_\beta$  denotes the covariant
derivative, the generalization of the partial derivative
$\partial/\partial x^\beta$ to curvilinear coordinates or curved
spaces. In coordinates, the covariant derivative of a vector field
$a^\mu$
acquires the form
\begin{equation}\label{covder}
\nabla_\alpha a^\mu=
 \frac{\partial a^\mu}{\partial
 x^\alpha}+\Gamma^\mu{}_{\alpha\beta}a^\beta.
\end{equation}
 The Christoffel symbol $\Gamma^\mu{}_{\alpha\beta}$ can be calculated
when the spatial dependence of the metric tensor is known
\begin{equation}\label{christoffel}
\Gamma^{\lambda}{}_{\mu\nu}
=\frac{1}{2}\gamma^{\lambda\sigma}\left(\frac{\partial
    \gamma_{\sigma\mu}}{\partial x^{\nu}}+\frac{\partial
    \gamma_{\sigma\nu}}{\partial x^{\mu}}-\frac{\partial
    \gamma_{\nu\mu}}{\partial x^{\sigma}}\right).
\end{equation}
It accounts for the fact that not only the vector components but also
the coordinate vectors themselves depend on position. Consequently,
the covariant derivative of a scalar field is equal to the partial
derivative. 
Details can be found in any book on general relativity,
e.g., \cite{wald,misner}.
In the present paper we will  avoid the application
of explicit coordinates, since our goal is to show the fundamental
geometrical structure experienced by light propagating through an 
inhomogeneous medium. Obviously, as soon as one is interested in
 a concrete example, one has to choose a system of coordinates. 

Maxwell's equations in the form (\ref{maxwell3d1}) and
(\ref{maxwell3d2}) have been introduced formally as an alternative  to
the standard form (\ref{maxwell3d}). They are obviously correct when
the standard metric is used, but one has to check if they remain valid
 in the case of alternative metric tensors.  The metric
tensors we consider here are connected to the flat
metric via a conformal transformation, i.e.
\begin{equation}\label{confmetric}
\gamma_{\mu\nu}=\Omega^2\delta_{\mu\nu}
\end{equation}
with $\Omega$ being some differentiable positive scalar-valued
field. 

When changing to a conformally equivalent metric it may become necessary
to introduce also transformed fields of the form
\begin{equation}
\tilde{D}_\alpha=\Omega^s D_\alpha
\end{equation}
where $s$ is an integer called the conformal weight. In order to
calculate it one has to use the explicit form of the covariant
derivative  (\ref{covder}) with the Christoffel symbol in the form
\begin{equation}\label{chrisconf}
\Gamma^{\alpha}{}_{\mu\nu}=\tilde{\nabla}_\nu(\mbox{ln}\, \Omega)
\delta^{\alpha}_{\mu}+\tilde{\nabla}_\mu(\mbox{ln}\,
\Omega)\delta^{\alpha}_{\nu}-
\tilde{\nabla}_\beta(\mbox{ln}\, \Omega)\delta^{\alpha\beta}\delta_{\mu\nu}.
\end{equation}
Eq.\ (\ref{maxwell3d2}) then acquires the form
\begin{eqnarray}
\tilde{\gamma}^{\alpha\beta}\tilde{\nabla}_{\alpha}\tilde{D}_{\beta}&=&
\Omega^{-2}\gamma^{\alpha\beta}(\partial_\alpha\Omega^sD_\beta
-\Omega^s \Gamma^\gamma_{\alpha\beta}D_\gamma)\\
\label{confweight}
&=&\Omega^{-2}((s-2+d)(\Omega^{s-1} D^{\beta}\partial_{\beta} \Omega) +
  \Omega^s\partial_{\alpha}D^{\alpha})
\end{eqnarray}
where $d$ is the dimension of the space,which in our case is 3.
 For eq.\ (\ref{maxwell3d2}) to be equivalent to the usual Maxwell
 equation the first term in (\ref{confweight}) has to vanish, leading
 to the result
\begin{equation}
s=-1
\end{equation}
which therefore implies
\begin{equation}\label{transd}
\tilde{D}_\alpha=\Omega^{-1}D_\alpha\qquad\mbox{and}
\qquad\tilde{D}^\alpha=\Omega^{-3}D^\alpha
\end{equation}
with an analogous relation for the $B$-field. Because the Christoffel
symbol is symmetric in the two lower indices we can rewrite eq.\ 
(\ref{maxwell3d1}) as
\begin{equation}\label{rot}
\frac{1}{\sqrt{\gamma}}e^{\alpha\beta\gamma}
\frac{\partial \tilde{E}_{\gamma}}{\partial x^{\beta}}
+\frac{\partial \tilde{B}^{\alpha}}{\partial t}=0
\end{equation}
and, taking into account that 
 $\sqrt{\gamma}=\Omega^3$, one gets 
\begin{equation}\label{transe}
\tilde{E}_\alpha=E_\alpha\qquad\mbox{and}
\qquad\tilde{E}^\alpha=\Omega^{-2}E^\alpha
\end{equation}
with the equivalent result for the $H$-field.
These results lead to the alternative constitutive equations
\begin{eqnarray}
\tilde{B}_\alpha=\Omega^{-1}B_\alpha=\Omega^{-1}{\mu\mu_0}
H_\alpha=\Omega^{-1}{\mu\mu_0} \tilde{H}_\alpha\\
\tilde{D}_\alpha=\Omega^{-1} D_\alpha=\Omega^{-1}\varepsilon\varepsilon_0
E_\alpha=\Omega^{-1}\varepsilon\varepsilon_0 \tilde{E}_\alpha,
\end{eqnarray}
valid in the transformed metric.
This allows us to define the permeabilities of the  system in the new
metric as
\begin{equation}\label{tildedef}
\tilde{\mu}=\frac{\mu}{\Omega}\quad\mbox{and}\quad\tilde{\varepsilon}=\frac{\varepsilon}{\Omega}.
\end{equation}
In order to arrive at the general form of the wave equation we
calculate as usual the curl of the first equation in
(\ref{maxwell3d2}) applying the covariant curl operator with
$\eta_{\alpha\beta\gamma}=\sqrt{\gamma}e_{\alpha\beta\gamma}$, and get
\begin{equation}\label{curlcurl}
\eta_{\alpha\rho\sigma}\tilde{\nabla}^\rho
\left(\eta^{\alpha\beta\gamma}\tilde{\nabla}_\beta \tilde{E}_\gamma+\frac{\partial
    \tilde{B}^\alpha}{\partial t}\right)=0.
\end{equation}
This leads to the equation
\begin{equation}\label{wave2}
\left(\delta^\beta_\rho\delta^\gamma_\sigma-\delta_\rho^\gamma\delta_\sigma^\beta\right)
\tilde{\nabla}^\rho\tilde{\nabla}_\beta \tilde{E}_\gamma+\frac{\partial}{\partial
  t}\eta_{\alpha\rho\sigma}\tilde{\nabla}^\rho
\tilde{B}^\alpha=0.
\end{equation}
Here we applied the fact that the covariant derivative of the modified
antisymmetric tensor $\eta^{\alpha\beta\gamma}$ vanishes. This can be
easily seen, taking into account that the covariant derivative of the
antisymmetric tensor $e^{\alpha\beta\gamma}$ does not vanish, but rather
leads to the result
\begin{equation}
\tilde{\nabla}_\rho e^{\alpha\beta\gamma}=\Gamma^{\sigma}_{\, \sigma\rho}
e^{\alpha\beta\gamma}=\frac{\partial \mbox{ln}\,\sqrt{\gamma}}{\partial x^\rho}
e^{\alpha\beta\gamma}
\end{equation}
where in the last step the well-known result for the contracted
Christoffel symbol has been applied \cite{wald}. 
Note the order of indices in the antisymmetric tensor:
\begin{equation}
\eta_{\alpha\rho\sigma}\tilde{\nabla}^\rho
\tilde{B}^\alpha=-\eta_{\sigma\rho\alpha}\tilde{\nabla}^\rho \tilde{B}^\alpha
\end{equation}
with the last term corresponding to the curl of $B^\alpha$. 
Thus we get for (\ref{wave2})
\begin{equation}
\tilde{\nabla}^\beta\nabla_\beta
\tilde{E}_\sigma-\tilde{\nabla}^\gamma\tilde{\nabla}_\sigma 
\tilde{E}_\gamma
-\frac{\partial}{\partial t}\eta_{\sigma\rho\alpha}\tilde{\nabla}^\rho
\tilde{B}=0.
\end{equation}
In order to be able to apply the divergence equation
 (\ref{maxwell3d2}),
 one has to
interchange the covariant derivatives in the second term. But, as one
easily sees from the explicit form in eq.\ 
(\ref{covder}), covariant derivatives  do not usually commute. The
commutator of covariant derivatives is connected to the curvature
of space, e.g., when applied to a vector the commutator gives 
\begin{equation}
[\tilde{\nabla}_{\alpha},\tilde{\nabla}_{\beta}]a^{\mu}=R^{\mu}{}_{\nu\alpha\beta}a^{\nu}
\end{equation}
with $R^{\mu}{}_{\nu\alpha\beta}$ being the Riemann curvature tensor
\cite{wald,misner}. Note that the conformal change of the metric
tensor may create a curvature, and the fact that the considered
space is flat in the standard metric does not imply that this
remains so after the transformation of the metric. 
In our case,the commutator gives 
\begin{equation}
[\tilde{\nabla}_{\alpha},\tilde{\nabla}_{\mu}]\tilde{E}^{\mu}= R_{\alpha\mu}\tilde{E}^{\mu}
\end{equation}
where 
\begin{equation}\label{ricci}
R_{\alpha\mu}=R^{\beta}{}_{\alpha\beta\mu}
\end{equation}
is the Ricci tensor. 
Finally, we get the wave equation in the form:
\begin{equation}\label{waveequation3d}
\tilde{\nabla}^\beta\tilde{\nabla}_\beta
\tilde{E}_\sigma+\tilde{\nabla}_\sigma(\tilde{E}_\gamma\tilde{\nabla}^\gamma\mbox{ln}\,
\tilde{\varepsilon}
)-\frac{\tilde{\mu}\tilde{\varepsilon}}{c^2}\frac{\partial^2}{\partial
  t^2}\tilde{E}_\sigma+R_{\alpha\mu}\tilde{E}^{\mu}+
(\tilde{\nabla}_\sigma \tilde{E}_\rho-\tilde{\nabla}_\rho
\tilde{E}_\sigma)
\tilde{\nabla}^\rho\mbox{ln}\,\tilde{\mu}=0.
\end{equation}
One easily notes that when the flat metric is applied, this is the
usual wave equation for light in an inhomogeneous medium,  e.g. \cite{bw}:
\begin{equation}\label{waveequation0}
\nabla^2{\bf E}+\frac{\mu\varepsilon}{c^2}\frac{\partial^2 {\bf
    E}}{\partial t^2}+(\mbox{grad ln }\mu)\times (\nabla\times {\bf
    E})+\mbox{grad}({\bf E}\cdot\mbox{grad ln }\varepsilon)=0.
\end{equation}

\section{Geometrical Optics}\label{geomoptics}
The physical content of the metric transformations becomes clear in
the framework of geometrical optics, i.e., in the case of monochromatic
light moving in a medium with the medium parameters not changing
significantly within one wavelength. This approximation allows us to
introduce the notion of light rays propagating through the
medium. The field can  then be  written in the form
\begin{equation}\label{field3d}
\tilde{E}_\sigma=\tilde{\cal E}_\sigma \exp(i (k_0{\cal S}-\omega t)).
\end{equation}
$\tilde{\cal E}_\sigma$ is the envelope of the wave; other fields like $\tilde{\cal
H}^\alpha$ or $\tilde{\cal D}^\alpha$ are defined likewise. $\omega$ is the
constant frequency and $k_0$ is defined as
$k_0=\omega/c=2\pi/\lambda_0$ with $\lambda_0$ being the vacuum wave
length. The product of $k_0$ and $\cal S$ -- the ``optical path'' --
is the spatial phase of the wave. Inserting (\ref{field3d}) into the wave
equation (\ref{waveequation3d}) and expanding in orders of $1/k_0$ gives:
\begin{eqnarray}\label{geowave}
0=&-&\tilde{\cal E}_\sigma \tilde{\nabla}_\beta{\cal S}\tilde{\nabla}^\beta{\cal S}
+\tilde{\mu}\tilde{\varepsilon}\tilde{\cal E}_\sigma\\ \nonumber
&+&\frac{i}{k_0}\left(2\nabla^\beta\tilde{\cal E}_\sigma\tilde{\nabla}_\beta{\cal S}+\tilde{\cal
  E}_\sigma
\tilde{\nabla}^\beta\tilde{\nabla}_\beta{\cal S}+
\tilde{\cal E}_\gamma\tilde{\nabla}_\sigma{\cal
  S}\tilde{\nabla}^\gamma\mbox{ln}\,(\tilde{\mu}\tilde{\varepsilon})-\tilde{\cal
  E}_\sigma\tilde{\nabla}_\rho{\cal S}\tilde{\nabla}^\rho\mbox{ln}\,\tilde{\mu}\right)
\\ \nonumber
&+&\frac{1}{k_0^2}\left(\tilde{\nabla}^\beta\tilde{\nabla}_\beta\tilde{\cal E}_\sigma+
\tilde{\nabla}^\gamma\mbox{ln}\,(\tilde{\mu}\tilde{\varepsilon})
\nabla_\sigma{\cal E}_\gamma
+\tilde{\nabla}^\beta\mbox{ln}\,\tilde{\mu}\tilde{\nabla}_\beta\tilde{\cal
  E}_\sigma-\tilde{\nabla}_\rho\tilde{\cal E}_\sigma\tilde{\nabla}^\rho\mbox{ln}\,\tilde{\mu}
+R_{\alpha\mu}\tilde{\cal E}^{\mu}\right).
\end{eqnarray}

One should not be confused by the fact that the expansion coefficient
has a dimension. The crucial point here is that two kinds of spatial
variation are present here: the fast variation of the phase and the
slow variation of the envelope. The 
clear separation of these scales is not a result of the calculations
but rather a condition for the geometrical approximation to be
valid. This validity is reflected in the size of $k_0$ which has to be
large enough for the three terms in (\ref{geowave}) to be
well-separated. 

Confining our considerations to the zeroth order contribution, we
get the equation
\begin{equation}\label{zeroth}
\nabla^\beta{\cal S}\nabla_\beta{\cal S}=\tilde{\mu}\tilde{\varepsilon}.
\end{equation}
In a homogeneous medium light rays follow straight lines, but in an
inhomogeneous one this is usually not the case. In the case of a
general metric the idea of the ``straightest line'' is represented by
the concept of a geodesic -- the line that ``curves as little as
possible'' \cite{wald} 
and that extremizes the distance  between two points. In
mathematical terms the geodesic is most easily
 described as the line that
``parallel transports its own tangent vector'', i.e., the 
covariant derivative of the tangent
vector field in the direction of the curve vanishes. 
Calling a tangent vector $t^\alpha$ we thus get the
condition 
\begin{equation}\label{geoddef}
t^\alpha\nabla_\alpha t^\beta=0
\end{equation}
for a curve to be a geodesic. 
In the case of propagating
light, the tangent vector is the velocity vector $\nabla^\alpha\cal S$,
 the contravariant
version of the wave vector $\nabla_\alpha{\cal S}$, which is the
gradient of the phase $\cal S$. 
Consequently, for the light ray to be a geodesic the 
condition
 \begin{equation}\label{geos}
\nabla^\alpha{\cal S}\nabla_\alpha \nabla^\beta{\cal S}=0
\end{equation}
must be fulfilled.
Note that $\cal S$ is a scalar and therefore its first derivative does
not depend on the metric and  two derivatives
applied to it commute. Eq.\ (\ref{geos}) can be rewritten in the form 
\begin{equation}
\frac{1}{2}\nabla^\beta\left(\nabla^\alpha{\cal S}\nabla_\alpha{\cal
    S}\right)=0.
\end{equation}
Since the expression in the brackets is a scalar, we end up with the
condition that  for the light trajectory to be a geodesic, 
$\nabla^\alpha{\cal S}\nabla_\alpha{\cal S}$ has to be constant. This
is equivalent to the condition that $\varepsilon\mu$ has to be
constant. This condition allows us to find the ``most natural'' metric
for a given medium. 
 Using (\ref{tildedef}) we conclude that
 the light ray is
 a geodesic in the metric
\begin{equation}\label{metric}
\bar{\gamma}_{\mu\nu}=\left(
\begin{array}{ccc}
n^2 & 0 & 0\\0 & n^2 & 0\\0 & 0 & n^2 \end{array}
\right),
\end{equation}
 i.e., one has to
 apply the
 conformal transformation with
\begin{equation}
\Omega^2=n^2.
\end{equation}
 In (\ref{metric}) we denoted the metric that allows us to represent
 the trajectories as
geodesics with a bar over the symbol. All objects corresponding to
this metric will be denoted by bars over the respective symbols,
whereas symbols without bars correspond to arbitrary metrics of the
form (\ref{confmetric}). Obviously, the metric tensor in
(\ref{metric}) may still be multiplied by a constant but our choice
 makes $\bar{\mu}$ and $\bar{\varepsilon}$ inverse. Note in particular
 that the product of $\bar{\mu}$ and $\bar{\varepsilon}$ corresponds to
 an effective index of refraction. Consequently, the metric introduced
 here leads to a constant index of refraction and thus to a system
 with a constant effective velocity of light.

The quantity $\bar{\mu}$ introduced here is equal to 
\begin{equation}\label{m}
m=\sqrt{\frac{\mu}{\varepsilon}}.
\end{equation}
 Together with the index of
refraction $n$, $m$ gives  an alternative description of the medium
properties. The interesting point is that $m$ drops out in the
calculations. Therefore it is possible to give a complete 
description of ray propagation  in an inhomogeneous medium in terms of
geodesics of a metric tensor. The quantity $m$ is sometimes called the
impedance \cite{scd} or the resistance \cite{birula} of the
medium. This is due the the fact that it has the units of Ohms. 

\section{Polarization transport}\label{partrans}
Since the electromagnetic field is not a scalar field, the knowledge
of the trajectory of a light ray -- the set of the points it passes
through -- does not contain complete information about the
propagation. One still needs to find out how the field strength
vectors are propagated along the ray. In order to address this
question we go back to eq.\ (\ref{geowave}), assume that eq.\
(\ref{zeroth}) is fulfilled and demand that even the first order
contribution vanishes. This leads to the equation
\begin{equation}\label{firstorder}
2\tilde\nabla^\beta\tilde{\cal E}_\sigma\tilde{\nabla}_\beta{\cal S}=-\tilde{\cal
  E}_\sigma\tilde{\nabla}^\beta\tilde{\nabla}_\beta{\cal S} - \tilde{\cal
  E}_\gamma\tilde{\nabla}_\sigma {\cal
  S}\tilde{\nabla}^\gamma\mbox{ln}\, (
\tilde{\varepsilon}\tilde{\mu})
+\tilde{\cal E}_\sigma
  \tilde{\nabla}_\rho{\cal S}\tilde{\nabla}^\rho\mbox{ln}\,\tilde{\mu}.
\end{equation}
 In analogy to 
(\ref{geoddef}) the vector $\cal E_\sigma$ is said to be parallel
transported, along the ray when the condition
\begin{equation}
\nabla^\beta{\cal E}_\sigma\nabla_\beta{\cal S}=0
\end{equation}
is fulfilled, the left-hand side corresponding to the directional derivative of
$\cal E_\sigma$ along the ray. In a more general sense the vector
$\cal E_\sigma$ is still said to be parallel transported if
\begin{equation}
\nabla^\beta{\cal E}_\sigma\nabla_\beta{\cal S}=\alpha{\cal E}_\sigma
\end{equation}
but in this case the length of the vector is not conserved during the
propagation. 
One sees
immediately that in order for the field vector to be parallel
transported the second term on the right hand side of
(\ref{firstorder}) has to vanish. This is equivalent to the condition
that $n^2$ is constant. Not surprisingly, this is exactly the same
condition as for the trajectory to be a geodesic. Consequently, in this
case we get the equation
\begin{equation}\label{parallel}
2\bar{\nabla}^\beta\bar{\cal E}_\sigma\bar{\nabla}_\beta{\cal S}=-\bar{\cal
  E}_\sigma\bar{\nabla}^\beta\bar{\nabla}_\beta{\cal S} +\bar{\cal E}_\sigma
  \bar{\nabla}_\rho{\cal S}\bar{\nabla}^\rho\mbox{ln}\,\bar{\mu}
\end{equation}
where we used the bar as introduced at the end of section
\ref{geomoptics}. 
The last term vanishes, if $\bar{\mu}$ is constant. This corresponds
to the condition that $\mu/n=\sqrt{\varepsilon/\mu}$ is constant in
the standard (or any other) metric,
i.e., $\varepsilon$ and $\mu$ differ only by a constant factor. This
means that $\bar{\mu}$ and $\bar{\varepsilon}$ are constant. In
this non-physical but interesting case, we get for the constitutive
equations
\begin{equation}
\bar{E}^\alpha=\bar{D}^\alpha\qquad \bar{B}^\alpha=\bar{H}^\alpha.
\end{equation} 
This means that the whole contribution of the medium iscontained in the
metric structure of space, the light is propagating as if it were
moving in empty but curved space. This result is not only valid in the
geometrical approximation, but even allows us to write the wave equation
in terms of a curved space as follows:
\begin{equation}\label{curvedwave}
\bar{\nabla}_{\rho}\bar{\nabla}_{\beta}\bar{E}_{\alpha}\gamma^{\rho\beta}+R_{\alpha\mu}
\bar{E}^{\mu}-\frac{\partial^2}{\partial
  t^2}\bar{E}_{\alpha}=0.
\end{equation}
This case corresponds to a constant value of the quantity $m$
introduced in (\ref{m}). 
In the general case of arbitrary $\mu$ and $\varepsilon$ we get for
the transformed wave equation:
\begin{equation}
\bar{\nabla}^\beta\bar{\nabla}_\beta \bar{E}_\sigma
-\bar{\mu}\bar{\varepsilon}\frac{\partial^2}{\partial t^2}\bar{E}_\sigma
+\bar{E}_\gamma\bar{\nabla}_\sigma\bar{\nabla}^\gamma \mbox{ln}\, \bar{\varepsilon}
+\bar{\nabla}_\rho \bar{E}_\sigma \bar{\nabla}^\rho\mbox{ln}\, \bar{\varepsilon}=0.
\end{equation}
As $\bar{\varepsilon}=1/m$, this equation shows that in the discussion
of the propagation of the field vectors in the general case both $n$
and $m$ have to be used. Since the geometrical description is based on
the index of refraction $n$ only, no complete geometric description is
possible in the general case.

With the field vectors being parallel
 transported along a geodesic,
the angles between the field vectors and between the fields and the
ray remain unchanged during the propagation along the ray. It is thus
sufficient to find these angles at one point on the ray in order to
know them for the whole trajectory. One can easily convince oneself that in
our approximation all the angles are right angles. 
This can be seen from the geometrical approximation of Maxwell's
equation by inserting the harmonic fields
(\ref{field3d}) into Maxwell's equations (\ref{maxwell3d1}) and 
(\ref{maxwell3d2}) and keeping only
the lowest order in $1/k_0$. We drop the tilde over the symbols when
 the general case is meant.  We thus get
\begin{equation}\label{geomax1}
\eta^{\alpha\beta\gamma}(\nabla_\beta{\cal S}){\cal H}_\gamma
+{\varepsilon\varepsilon_0}{\cal E}^\alpha=0\qquad\mbox{and}\qquad
\eta^{\alpha\beta\gamma}(\nabla_\beta{\cal S}){\cal E}_\gamma
-{\mu\mu_0}{\cal H}^\alpha=0
\end{equation}
and
\begin{equation}\label{geomax2}
{\cal D}^\alpha\nabla_\alpha{\cal S}=0\qquad\mbox{and}\qquad
{\cal B}^\alpha\nabla_\alpha{\cal S}=0.
\end{equation}
The last two equations show that the field vectors are orthogonal to
the ray. In order to show that they are orthogonal to each other one
has to multiply the first equation of (\ref{geomax1}) by 
${\cal H}_\alpha$ which leads to 
\begin{equation}
{\varepsilon\varepsilon_0}{\cal E}^\alpha{\cal H}_\alpha=0
\end{equation}
where the antisymmetry of the tensor $\eta^{\alpha\beta\gamma}$ has to
be taken into account. Finally, contracting the first equation of 
(\ref{geomax1}) with ${\cal E}_\alpha$ and the second one with  
${\cal H}_\alpha$ we arrive at 
\begin{equation}\label{ee}
{\varepsilon\varepsilon_0}{\cal E}^\alpha{\cal E}_\alpha=
{\mu\mu_0}{\cal H}^\alpha{\cal H}_\alpha.
\end{equation}
Note that the terms in  eq.\ (\ref{ee}) are -- up to a constant --  
the electric
and magnetic energy densities, respectively, showing that both
densities are equal. The complete energy density of either field
 can  be written in the form 
\begin{eqnarray}\label{we}
{w}_e&=&\frac{1}{4}{\cal E}_\alpha{\cal
    D}^\alpha=\frac{\varepsilon\varepsilon_0}{4}{\cal E}_\alpha{\cal
    E}^\alpha,\\ \label{wm}
{w}_m&=&\frac{1}{4}{\cal H}_\alpha{\cal
    B}^\alpha=\frac{\mu\mu_0}{4}{\cal H}_\alpha{\cal
    H}^\alpha.
\end{eqnarray}

The transformation of the fields to the new metric as described in
(\ref{transd}) and (\ref{transe}) leaves the direction of the field vectors
unchanged. Consequently, the result that they are orthogonal to each
other and to the ray does not depend on the metric. But in the general
metric the light ray is not a geodesic and consequently the vectors
parallel transported along the ray will point in another direction
than the physical ones. Thus what makes the metric (\ref{metric}) special
 is the
fact that the parallel transported and the physical field vectors coincide
along the whole ray. 

Note that the  energy densities, like the Poynting vector defined
later, are time-averaged quantities with the fast
temporal variation being disregarded. 
 It should be
mentioned that the numerical values of the energy densities (\ref{we})
and (\ref{wm})
depend on the metric chosen. This is due to the fact that the
physical quantity involved is the integral of the density  over some volume,
but as the line element is transformed according to
\begin{equation}\label{linetrans}
\widetilde{dx}^\alpha=\frac{1}{\Omega}dx^\alpha
\end{equation}
one has to change the density as well in order to keep the integral
constant. Eq.\ (\ref{ee}) suggests that we 
 introduce the auxiliary fields
\begin{equation}\label{km}
{\cal K}_\sigma=\sqrt{\varepsilon\varepsilon_0}{\cal
  E}_\sigma\qquad\mbox{and}\qquad{\cal
  M}_\sigma=\sqrt{{\mu\mu_0}}{\cal
  H}_\sigma.
\end{equation}
These vectors have equal absolute value and 
fulfill the simplified equation 
\begin{equation}\label{park}
2\nabla^\beta{\cal K}_\sigma\nabla_\beta{\cal S}=-{
  \cal K}_\sigma\nabla^\beta\nabla_\beta{\cal S}.
\end{equation}

Eqs.\ (\ref{parallel}) and (\ref{park}) do not only describe the
 propagation of field vectors along the ray but they are  also the key
 to 
 understanding  the propagation of the field intensity. 
 The intensity $I$ of a light ray is
defined as the absolute value of the Poynting vector
\begin{equation}\label{Poynting}
S^\alpha=\frac{1}{2}\eta^{\alpha\beta\gamma}{\cal E}_\beta{\cal H}_\gamma
\end{equation}
and it is easily shown that it can be written in the form 
\begin{equation}\label{intdef}
I=2\frac{c}{\sqrt{\varepsilon\mu}}w_e.
\end{equation}
Inserting (\ref{zeroth}) into (\ref{intdef}) and taking into
account that $\bar{\varepsilon}\bar{\mu}=1$ we get for the
intensity transport the
result
\begin{eqnarray}
\bar{\nabla}^\beta{\cal S}\bar{\nabla}_\beta \bar{I}&=&
\frac{1}{2}\sqrt{\frac{\varepsilon_0}{\mu_0}}
\bar{\nabla}^\beta{\cal
  S}\bar{\nabla}_\beta(\bar{\varepsilon}\bar{{\cal E}}_\sigma \bar{{\cal
  E}}^\sigma)\\
&=&\frac{1}{2}\sqrt{\frac{\varepsilon_0}{\mu_0}}
(2\bar{\varepsilon}\bar{{\cal E}}^\sigma \bar{\nabla}^\beta{\cal S}\bar{\nabla}_\beta\bar{{\cal
  E}}_\sigma+\bar{{\cal E}}^\sigma\bar{\cal E}_\sigma\bar{\nabla}^\beta{\cal
  S}\bar{\nabla}_\beta \bar{\varepsilon})\\
&=&-\frac{1}{2}\sqrt{\frac{\varepsilon_0}{\mu_0}}[\bar{\varepsilon}\bar{\cal E}^\sigma \bar{\cal E}_\sigma
  (\bar{\nabla}_\beta\bar{\nabla}^\beta{\cal S}+\bar{\nabla}^\beta{\cal S}\bar{\nabla}_\beta
  \mbox{ln}\,{\bar{\varepsilon}}
 +\bar{\nabla}^\beta{\cal S}\bar{\nabla}_\beta
  \mbox{ln}\, {\bar{\mu}})]\\ \label{iprop}
&=&-\bar{I}\bar{\nabla}_\beta\bar{\nabla}^\beta{\cal S}.
\end{eqnarray}
The right-hand-side  of (\ref{iprop}) describes the change of the
electromagnetic intensity along the ray. Due to the fact that
$\bar{\mu}$ and $\bar{\varepsilon}$ are inverse, the terms containing
these quantities cancel and the final result does not depend on their
actual values. Consequently, light propagation in the geometrical
approximation does not crucially depend on 
whether $\bar{\mu}$ and $\bar{\varepsilon}$ are constant or not, and
neither the trajectory nor the energy flow are dependent on
that. Clearly, the change of the fields $E^\alpha$ and $H^\alpha$
along the ray does explicitly depend on the variation of 
$\bar{\mu}$ and $\bar{\varepsilon}$, respectively.
 The purpose of the last term in eq.\ (\ref{parallel}) is, in a sense,
 to adjust the field in a way that, despite the changing permeabilities,
 its intensity flow is half of (\ref{iprop}). As has
 been pointed out before the product of $\bar{\mu}$ and
 $\bar{\varepsilon}$ can be seen as an effective index of
 refraction. Consequently, the mathematical framework of the changed
 metric leads to a constant velocity of light and therefore 
corresponds to an empty but possibly curved space, for all
 physical processes where only the index of refraction is involved.

The result  in eq.\ (\ref{iprop}) 
describing the change in intensity along the ray
sheds light on the wave structure of
the discussed set up. This can be easily understood if one looks not
just
 at one trajectory but rather considers  a thin but finite bundle of
trajectories. Clearly, the distance between the trajectories may 
depend on the curve parameter. In the homogeneous case, however,
all light rays are straight lines, and consequently the dependence is
linear making the  second derivative of the distance vanish. One
can easily show that in this case the Laplacian of the phase
$\partial^\alpha\partial_\alpha{\cal S}$ is proportional to the mean
curvature, the arithmetic mean  of the two principal curvatures, of the
wave front. The formula for the change of the element of surface area
spanned by a bundle of straight lines is given in books on elementary
differential geometry (see e.g. \cite{oneill}) where the term
``parallel surfaces'' is used to denote the system of  successive
wave fronts. Consequently, the change in intensity is governed by the
curvature of the wave fronts, whereas the value of the intensity is
inversely proportional to the area of the cross section of the bundle.
In the inhomogeneous case the right-hand side of eq.\ (\ref{park})
 cannot be interpreted
in this simple way however because one has to add a term containing a
Christoffel symbol to the simple Laplace expression. The additional
term can be traced  to the fact that  in the inhomogeneous
medium geodesics do not continue straight on, but rather curve through
the 
dependence on the structure of the medium -- in
general in a different way at different points. Usually, the distance
between these geodesics does not depend linearly on the curve
parameter, i.e., the second derivative of the distance does not
vanish. This effect is known as the geodesic deviation and is
a characteristic of a curved space. Consequently, the changed law of
intensity variation can be seen as a sign of the fact that the
inhomogeneity of the medium creates an effectively curved space. 

More
sophisticated mathematical methods should allow for a discussion of
curved surfaces in a curved space. In this way (\ref{iprop})
might turn out to be the curved-space analogue of the mean curvature.

\section{The Riemann-Silberstein vector in inhomogeneous media}\label{rssec}
By analogy with the fields $\cal M_\alpha$ and $\cal K_\alpha$
in eq.\ (\ref{km}) one can generally
introduce the fields
\begin{equation}
K_\gamma=\frac{D_\gamma}{\sqrt{\varepsilon\varepsilon_0}}={E_\gamma}{\sqrt{\varepsilon\varepsilon_0}}
\end{equation}
and 
\begin{equation} 
M_\gamma=\frac{B_\gamma}{\sqrt{\mu\mu_0}}
={H_\gamma}{\sqrt{\mu\mu_0}}.
\end{equation}
These fields can be seen to be the real and imaginary parts of the
Riemann-Silberstein vectors
\begin{equation}\label{rs}
(F_\pm)_\alpha=\frac{1}{\sqrt{2}}(K_\alpha\pm iM_\alpha).
\end{equation}
They allow for an alternative description of electrodynamics
with the complete Maxwell equations in vacuum taking the form
\begin{equation}\label{sileq}
i\frac{\partial {\bf F}_\pm}{\partial t}=\pm c\nabla\times{\bf F}_\pm
\qquad{\rm and}\qquad\nabla\cdot{\bf F}_\pm=0.
\end{equation}
Clearly, only one of the fields is needed for a complete description of
electrodynamics. 
When used to describe  photons, the Riemann-Silberstein vector has to
be a correctly defined wave function, i.e., it may only include positive
frequencies. In this case both 
 vectors, $(F_+)_\alpha$ and  $(F_-)_\alpha$, have to be used, 
 corresponding to
 light which is left- and right-circular polarized,
respectively, i.e., they then correspond to positive and negative helicity
states  \cite{birula}. 
In an inhomogeneous medium the field equations for the
Riemann-Silberstein field become considerably more complicated than eq.\
(\ref{sileq}). Written in the general form with an arbitrary diagonal
metric they attain the form
\begin{eqnarray}\label{rotf}
i\frac{\partial F_-^\alpha}{\partial t}+
\frac{c\eta^{\alpha\beta\gamma}}{\sqrt{\mu\varepsilon}}\left(
\nabla_\beta (F_-)_\gamma+\frac{1}{2}\left((F_+)_\gamma
\nabla_\beta\mbox{ln}\,\sqrt{\frac{\mu}{\varepsilon}}+(F_-)_\gamma\nabla_\beta\mbox{ln}\,
\frac{1}{\sqrt{\mu\varepsilon}}\right)\right)&=&0\\ \nonumber
i\frac{\partial F_+^\alpha}{\partial t}-
\frac{c\eta^{\alpha\beta\gamma}}{\sqrt{\mu\varepsilon}}\left(
\nabla_\beta (F_+)_\gamma+\frac{1}{2}\left((F_-)_\gamma
\nabla_\beta\mbox{ln}\,\sqrt{\frac{\mu}{\varepsilon}}+(F_+)_\gamma\nabla_\beta\mbox{ln}\,
\frac{1}{\sqrt{\mu\varepsilon}}\right)\right)&=&0
\end{eqnarray}
and
\begin{eqnarray}\label{divf}
\nabla_\alpha
F_+^\alpha=\frac{1}{2}\left(F_+^\alpha\nabla_\alpha\mbox{ln}\,\frac{1}
{\sqrt{\varepsilon\mu}}+F_-^\alpha\nabla_\alpha\mbox{ln}\,
\sqrt{\frac{\mu}{\varepsilon}}\right)\\ \nonumber
\nabla_\alpha
F_-^\alpha=\frac{1}{2}\left(F_-^\alpha\nabla_\alpha\mbox{ln}\,\frac{1}
{\sqrt{\varepsilon\mu}}+F_+^\alpha\nabla_\alpha\mbox{ln}\,
\sqrt{\frac{\mu}{\varepsilon}}\right).
\end{eqnarray}
Note that the two helicity states usually couple, i.e., helicity is
not a constant of motion due to a term proportional to the derivative
of $\mbox{ln}\,\sqrt{\mu/\varepsilon}$. In the non-physical case discussed
before, i.e., when 
$\varepsilon$ and $\mu$ are proportional, the equations obviously decouple
and one gets for the wave equation in the transformed metric 
the same form as in ({\ref{curvedwave}):
\begin{equation}
-\frac{1}{c^2}
\frac{\partial^2(\bar{F}_{-})_{\sigma}}{\partial t^2}+R_{\sigma\mu}\bar{F}_-^\mu+
\bar{\nabla}^\beta\bar{\nabla}_\beta (\bar{F}_{-})_{\sigma}=0.
\end{equation}
As has been emphasized before, this choice of parameters allows us to
 describe
 light in an inhomogeneous medium exactly as if
it were propagating in a curved space. In this case all polarization
states are well defined and conserved. When as the ratio of
$\varepsilon$ and $\mu$ becomes variable, the equations (\ref{rotf})
and (\ref{divf})
 couple  again. But still, in the geometrical approximation both
fields propagate independently, because the wave equation takes the form:
\begin{equation}\label{fwave}
\frac{\varepsilon\mu}{c^2}\frac{\partial^2F_{-\sigma}}{\partial t^2}-
\nabla^\beta\nabla_\beta F_{-\sigma}+\nabla_\rho F_{-\sigma}\nabla^\rho
\mbox{ln}\,\frac{1}{\varepsilon\mu}-\nabla_\sigma
F^\gamma_-\nabla_\gamma\mbox{ln}\,\frac{1}{\varepsilon\mu}=0.
\end{equation}
Here all terms have been neglected which in the geometrical
 approximation  contribute
 to the highest order in $1/k_0$ 
 only, i.e., terms not containing any derivative of
 $F^\alpha$. Note that in this approximation the two 
 helicity states decouple in the wave equation in all metrics. In the
 case where 
 $\varepsilon$ and $\mu$ are
 inverse the two last terms on the right-hand side of (\ref{fwave}) vanish,
 leaving us with a standard wave equation expressed in terms of
 covariant derivatives. The discussion of geometrical optics in this
 case is then fully equivalent to that given in Section
 \ref{partrans}. Consequently, if the metric (\ref{metric}) is
 applied, the complex vector $\cal F^\alpha$ -- defined analogously to
 $\cal E^\alpha$ -- is parallel transported
 along the ray, which is a geodesic of the introduced metric. This
 means that the new metric allows to introduce well-defined
 polarization states which are propagated along the ray. In
 particular, helicity is a conserved quantity. Thus, if the
 Riemann-Silberstein vector had to be used to describe the wave
 function of a photon, the extension of this concept to inhomogeneous
 media should be discussed in the framework of quantum optics in a
 curved space.

\section{Summary}
In the present paper a general geometrical description of
electromagnetic  phenomena
 in inhomogeneous dielectric media has been given with the emphasis
being on light propagation in the approximation of geometrical optics. 
To find the conceptually simplest description we introduced  a
variation of the spatial metric allowing for all metric tensors
connected to the unit matrix by a conformal transformation. The
equations for the fields and for the propagation
of light rays can easily  be rewritten in this general geometry when
care is taken to  transform correctly the fields and the material parameters. 
It turns out that when the metric tensor is equal to the unit matrix
multiplied by $n^2$ -- the square of the index of refraction -- 
the equations become particularly simple:
light rays are geodesics of the metric and the field vectors are
parallel transported along the ray. Even the wave equation simplifies
significantly. As long as $\mu$ and $\varepsilon$ in the standard
metric are independent quantities the system in the transformed metric
corresponds to a curved space filled with a dielectric medium, but with
the transformed quantities $\bar{\mu}$ and $\bar{\varepsilon}$
having a constant product. This property leads to  significant
simplifications in the description of light propagation. In the
special case where $\mu$ and $\varepsilon$ differ only by a constant
factor the situation becomes even simpler with the transformed
qualities becoming constants. In this -- non-physical -- case the
transformation fully eliminates the medium and all electromagnetic
effects appear as if they were taking place in an empty but curved
space. Whereas the wave equations contain significantly more terms
when an effective medium is present, there are only slight differences
between the two cases in the approximation of geometrical
optics. The results there depend mostly on the behavior of the
effective index of refraction which becomes constant. 
 
The results presented here contribute to a better fundamental understanding of
electromagnetism in dielectric media and show  the power
of geometrical concepts in classical fields of physics. 
In particular the mathematical techniques of the general theory of
relativity turn out to be  well-suited for calculations in
inhomogeneous media. 
In a
forthcoming paper the  presented
results will be extended to electromagnetic fields in 
moving inhomogeneous media, thus supplementing the calculations
presented in our papers on light in moving media. A tempting problem
might be a geometric theory of anisotropic media which will require 
more sophisticated mathematical methods than those presented here
including the introduction of a more general space than the Riemannian
space. Another mathematical challange would be a
discussion of the wave fronts in the spirit of differential geometry
of curved surfaces in a curved three-dimensional space. 
Furthermore  our results show that the correctly defined 
Riemann-Silberstein vector keeps  a well-defined helicity during its
propagation through the medium. This might trigger further investigations in
the search for a photon wavefunction in a dielectric medium. 

\begin{acknowledgements}
The author thanks Ulf Leonhardt at the University of St.\ Andrews for
valuable discussions and Stig Stenholm at the Royal Institute of
Technology in Stockholm for his continuous support. Thanks to Iwo
Bia{\l}ynicki-Birula for valuable comments. 
\end{acknowledgements}


\begin{thebibliography}{99}
\bibitem{movmed1}
U. Leonhardt \& P.\ Piwnicki, Phys. Rev. A {\bf 60}, 4301 (1999)
\bibitem{sound}  
Matt Visser, Class. Quant. Grav.{\bf  15}, 1767 (1998); 
L. J. Garay, J. R. Anglin, J. I. Cirac,  and P. Zoller, 
Phys. Rev. A {\bf 63}, 023611 (2001); Phys. Rev. Lett. {\bf 85}, 4643 (2000)
\bibitem{bort} E. Bortolotti, Rend. R. Acc. Naz. Linc., 6a, {\bf 4}, (1926)
\bibitem{rytov} S.M. Rytov, Compt. Rend. (Doklady) Acad. Sci. URSS,
  {\bf 18} (1938), 263
\bibitem{klinekay} M. Kline \& I.W. Kay, {\it Electromagnetic Theory
  and Geometrical Optics}, Interscience Publishers, 1965
\bibitem{scd} S. Solimeno, B. Crosignani, and P. DiPorto, {\it Guidin,
    Diffraction, and Confinement of Optical Radiation}, Academic
  Press, 1986 
\bibitem{weber} H. Weber, {\it Die partiellen Differentialgleichungen
    der mathematischen Physik nach Riemanns Vorlesungen}, (Friedrich
  Vieweg und Sohn, Brauenschweig, 1901)
\bibitem{silberstein} L. Silberstein, Ann. d. Phys. {\bf 22}, 579
  (1907); {\bf 24}, 783 (1907)
\bibitem{birula} Iwo Bia{\l}ynicki-Birula, in {\it Coherence and
    Quantum Optics VII},
edited by J. H. Eberly, L. Mandel, and E. Wolf, (Plenum, New York 1996), p. 313
\bibitem{ll2100ex} L.D.\ Landau and E.M.\ Lifshitz, {\it Classical Theory of
    Fields}, Pergamon, 1980; Problem in \S 90
\bibitem{wald} R. M.\ Wald, {\it General Relativity}
(The University of Chicago Press, Chicago and London, 1984).
\bibitem{misner} C. M. Misner, K. S. Thorne, and J. A. Wheeler,
 {\it Gravitation} (W.H. Freeman and Company, New York, 1973)
\bibitem{waldd} See Appendix D of Ref.\ \cite{wald} 
\bibitem{bw} M. Born and  E. Wolf, {\it Principles of Optics},
    (Pergamon, Oxford, 1980).
\bibitem{oneill} B. O'Neill, {\it Elementary Differential Geometry},
  (Academic Press, Orlando, 1966)


\end{thebibliography}
\end{document}